\documentclass[aps,prl,twocolumn,superscriptaddress,showpacs]{revtex4}

\usepackage{graphics}

\begin{document}

\title{New Type of Magneto-Rotational Instability in Cylindrical
Taylor-Couette Flow}

\author{Rainer Hollerbach}
\affiliation{Isaac Newton Institute for Mathematical Sciences,
Cambridge, CB3 0EH, United Kingdom}
\affiliation{Department of Mathematics, University of Glasgow,
Glasgow, G12 8QW, United Kingdom}

\author{G\"unther R\"udiger}
\affiliation{Isaac Newton Institute for Mathematical Sciences,
Cambridge, CB3 0EH, United Kingdom}
\affiliation{Astrophysikalisches Institut Potsdam, An der Sternwarte 16,
D-14482 Potsdam, Germany}

\date{\today}

\begin{abstract}
We study the stability of cylindrical Taylor-Couette flow in the presence
of combined axial and azimuthal magnetic fields, and show that adding an
azimuthal field profoundly alters the previous results for purely axial
fields.  For small magnetic Prandtl numbers $Pm$, the critical Reynolds
number $Re_c$ for the onset of the magneto-rotational instability becomes
independent of $Pm$, whereas for purely axial fields it scales as $Pm^{-1}$.
For typical liquid metals, $Re_c$ is then reduced by several orders of
magnitude, enough that this new design should succeed in realizing this
instability in the laboratory.
\end{abstract}


\maketitle

The magneto-rotational instability (MRI) is one of the most important
processes in astrophysics, with applications ranging from accretion disks
to entire galaxies \cite{r1,r2}.  There is therefore great interest
in trying to study this instability in the laboratory
\cite{r3,r4,r5,r6,r7,r8,r9,r10}.  These
experiments are complicated by the extremely small magnetic Prandtl
numbers of liquid metals, which necessitate very large rotation rates.
Here we report on a new type of magneto-rotational instability, that
operates even at infinitesimal Prandtl number, and correspondingly at
far lower rotation rates.  This new instability should be much
easier to obtain experimentally, and may also have considerable
astrophysical implications.
 
The magneto-rotational instability is a mechanism for transporting
angular momentum.  Consider, for example, an accretion disk around a
newly forming star.  In order for material to fall inward, it must
transfer its angular momentum to material further out.  The
difficulty is how to accomplish this.  In particular, a Keplerian
angular velocity profile, for which $\Omega\sim r^{-3/2}$, is known
to be hydrodynamically stable (at least linearly), by the familiar
Rayleigh criterion.  Purely viscous coupling in a laminar flow is many
orders of magnitude too small though; if this were the angular momentum
transport mechanism, stars would take so long to form that none
would yet exist today, some 14 billion years after the Big Bang.
 
This conundrum was solved by Balbus \& Hawley \cite{r1}, who showed
that such a Keplerian profile may be hydrodynamically stable, but is
nevertheless magnetohydrodynamically unstable.  The addition of a
magnetic field opens up a new way of coupling fluid parcels,
namely via the magnetic tension in the field lines, and thereby allows
for a new instability, the magneto-rotational instability, that has no
analog in the purely hydrodynamic problem.  Coupling the fluid in this
way is then so much more efficient at transporting angular momentum
that this mechanism can indeed yield accretion rates more in line with
those observed.  Something as basic as the time it takes for a star to
form is thus magnetically controlled.
 
Subsequent to this work, it was soon realized that this instability had
actually been discovered decades earlier, by Velikhov \cite{r11}, who had
not considered it in an astrophysical context though.  Instead, he had
viewed it simply as the magnetohydrodynamic extension of the so-called
Taylor-Couette problem, consisting of the flow between differentially
rotating cylinders.  See for example \cite{r12} for a review of this
problem, one of the oldest in fluid dynamics.
 
Once the connection is made to the classical Taylor-Couette problem,
one immediately recognizes that this should be a way to study this
instability in the lab: just take the fluid to be a liquid metal,
and apply a magnetic field along the cylinders.  This simple design
is the basis of most of the MRI experiments proposed to date
\cite{r3,r4,r5,r6,r7,r8}.  (Sisan {\it et al.}\ \cite{r9} claim to have
obtained the magneto-rotational instability already, in spherical rather
than cylindrical geometry.  However, there are enough other instabilities
that can arise in this configuration, e.g. \cite{r13}, that their results
are also open to other interpretations.  Indeed, the basic state from
which their instability arises is already fully turbulent, indicating
that whatever their instability may be, it is certainly not the first
instability to set in.)

As simple as it sounds, there is unfortunately also one considerable
difficulty associated with these experiments, namely that the rotation
rates of the inner and outer cylinders must be enormous.  The problem
is that the relevant parameter is not so much the hydrodynamic
Reynolds number $Re=\Omega_i r_i^2/\nu$, but rather the magnetic
Reynolds number $Rm=\Omega_i r_i^2/\eta$, where $\nu$ is the viscosity
and $\eta$ the magnetic diffusivity.  In its traditional form, the
magneto-rotational instability
sets in when $Rm=O(10)$.  $Re$ is then given by $Rm/Pm$, where $Pm=
\nu/\eta$ is the magnetic Prandtl number, a material property of the
fluid.  Typical values are $\sim$$10^{-5}$ for liquid sodium,
$\sim$$10^{-6}$ for gallium, and $\sim$$10^{-7}$ for mercury.  $Re$ must
therefore be at least $10^6$, and possibly larger still, depending on
the particular liquid metal one intends to use.
 
Such large values can be reached in the lab; taking the inner cylinder
radius $r_i=10$ cm, say, one finds that rotation rates of around 10 Hz
are required -- large, but achievable.  The difficulty lies
elsewhere, namely in the ends that would necessarily be present
in any real experiment.  Due to the Taylor-Proudman theorem, stating
that in rapidly rotating systems the flow will tend to align itself along
the axis of rotation, end-effects become increasingly important, until
at $Re=O(10^6)$ the flow is controlled almost entirely by the end-plates
\cite{r14}.  That is, while it is possible to achieve such large rotation
rates, the flow will look nothing like the idealized infinite-cylinder
flow on which all of the theoretical analysis is based.
 
A new approach is therefore needed if this instability is to be
studied in the laboratory.  We propose the following:  instead of
imposing only a uniform axial field $B_z=B_0$, additionally also
impose an azimuthal field $B_\phi=\beta B_0 (r_i/r)$, where $\beta$
is a nondimensional measure of the relative magnitudes of $B_\phi$
and $B_z$, and the dimensional quantity $B_0$ will be incorporated
into the Hartmann number below.  Such a $B_\phi$ field
can be generated just as easily as a $B_z$ field, by running a
current-carrying wire down the axis of the inner cylinder (suitably
insulated from the fluid, of course).  Note also that with neither
$B_z$ nor $B_\phi$ maintained by currents within the fluid itself, the
possibility of magnetic instabilities is excluded {\it a priori}.  The
only source of energy to drive an instability is the imposed
differential rotation; the magnetic field merely acts as a catalyst.

Given this basic state consisting of these externally imposed magnetic
fields, as well as the differential rotation profile $\Omega(r)$ imposed
by the rotation rates $\Omega_i$ and $\Omega_o$ of the two cylinders, we
linearize the governing equations about it, and look for axisymmetric
disturbances.  These are known to be preferred over non-axisymmetric ones
for the classical $\beta=0$ MRI, and are therefore the appropriate starting
point here as well.  The perturbation flow and field may then be expressed
as
\[{\bf u}=v{\bf\hat e}_\phi + \nabla\times(\psi{\bf\hat e}_\phi),
  \qquad
  {\bf b}=b{\bf\hat e}_\phi + \nabla\times( a  {\bf\hat e}_\phi).\]
Taking the $z$ and $t$ dependence to be $\exp(ikz+\gamma t)$,
the perturbation equations become
\[Re\,\gamma\,v=D^2v + Re\,ik\,r^{-1}(r^2\Omega)'\,\psi
      + H\!a^2\,ik\,b,\]
\[Re\,\gamma\,D^2\psi=D^4\psi-Re\,2ik\,\Omega\,v
   +H\!a^2\,ik(D^2a+2\beta r^{-2}b),\]
\[Pm\,Re\,\gamma\, b=D^2b-Pm\,Re\,ik\,\Omega'r\,a
    +ik\,v-2ik\,\beta r^{-2}\psi,\]
\[Pm\,Re\,\gamma\, a=D^2a + ik\,\psi,\]
where $D^2=\nabla^2-1/r^2$, and the primes denote $d/dr$.
 
Length has been scaled by $r_i$, time by $\Omega_i^{-1}$, $\Omega$ by
$\Omega_i$, $\bf u$ by $\eta/r_i$, and ${\bf B}_0$ and $\bf b$ by $B_0$.
The various nondimensional parameters are then: (a) the magnetic Prandtl
number $Pm=\nu/\eta$ already mentioned above, (b) the ratio $\hat\mu=
\Omega_o/\Omega_i$ (this enters into the details of $\Omega(r)=c_1+c_2/r^2$)
and the Reynolds number $Re=\Omega_i r_i^2/\nu$, measuring the relative and
absolute rotation rates of the two cylinders, (c) the parameter $\beta$ and
the Hartmann number $H\!a=B_0r_i/\sqrt{\rho\mu\eta\nu}$ ($\rho$ is the fluid's
density, $\mu$ the magnetic permeability), measuring the relative and absolute
magnitudes of the imposed magnetic fields, and (d) the radius ratio $r_i/r_o$,
which we fixed at $1/2$.
 
The radial structure of $v$, $\psi$, $b$ and $a$ was expanded in terms
of Chebyshev polynomials, typically up to $N=40-80$.  These equations and
associated boundary conditions (no-slip for $\bf u$, insulating for $\bf b$)
then reduce to a large ($4N$ by $4N$) matrix eigenvalue problem, with the
eigenvalue being the growth or decay rate $\gamma$ of the given mode.
Note also that this numerical implementation is very different from that
of \cite{r4}, in which the individual components of $\bf u$ and $\bf b$ were
used, and discretized in $r$ by finite differencing.  Both codes yielded
identical results though in every instance where we benchmarked one against
the other.
 
The following sequence of calculations was then carried out.  First, we fixed
$Pm$, $\beta$ and $\hat\mu$, and scanned through a range of values of $Re$,
$H\!a$ and $k$, in each case determining whether the given modes grow or decay.
We thus found the smallest value of $Re$, and the corresponding $H\!a$
and $k$, that yields a marginally stable solution, one having 
Re$(\gamma)=0$ (which typically involved solving the basic
eigenvalue problem for several thousand combinations of $Re$, $H\!a$ and $k$).
By repeating this entire procedure for different values of $Pm$, $\beta$ and
$\hat\mu$, we obtained the results in Fig.\ 1.

The dotted curve to the left of $\hat\mu=0.25$ is a purely hydrodynamic
instability, namely the onset of Taylor vortices, e.g. \cite{r12}.  As we
approach $\hat\mu=0.25$ though, we note that $Re_c\to\infty$.  This
critical value $\hat\mu=0.25$ ($(r_i/r_o)^2$, in general) is precisely
the so-called Rayleigh line, beyond which the flow is hydrodynamically
stable, because the angular momentum increases outward.

We are more interested therefore in the behavior to the right of
$\hat\mu=0.25$, where we anticipate that the inclusion of magnetic
effects will yield the MRI.  We begin with the two curves $\beta=0$,
the MRI as it has been considered to date \cite{r10}.  We see how $Re_c$
rises very steeply from the previous nonmagnetic results to the left
of $\hat\mu=0.25$, and then scales as $Pm^{-1}$, exactly as described
above, and in perfect agreement with \cite{r4}.  And again, because $Pm$
is so small, these values end up being too large for the experiment to
succeed \cite{r14}.
 
However, turning next to the results for non-zero $\beta$, we see
that $Re_c$ is dramatically reduced, over a range of $\hat\mu$
extending increasingly far beyond the Rayleigh line.  For example, if we
focus on how far we can go before $Re_c=10^4$, say, we obtain $\hat\mu=0.253$,
0.264, 0.292 and 0.308 for $\beta=1$, 2, 4 and 8, respectively.
Furthermore, within these $\hat\mu$ ranges $Re_c$ is independent of $Pm$,
very different from the previous $Pm^{-1}$ scaling.  Indeed, within these
ranges one can set $Pm=0$ and still obtain exactly the same solutions.  This
$Pm=0$ limit was considered before by Chandrasekhar \cite{r15}, but for
axial fields only, in which case there are no instabilities to the right
of the Rayleigh line.

\begin{figure}
\includegraphics{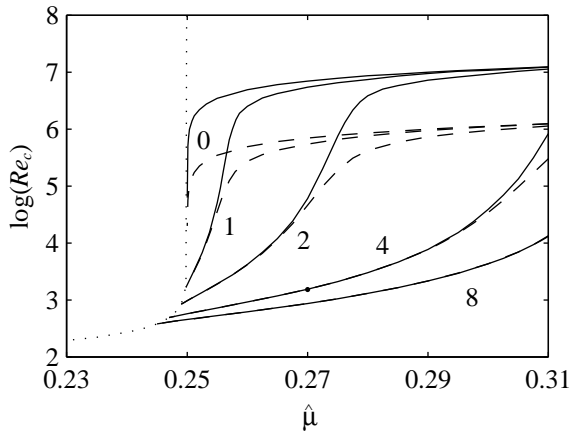}
\caption{The critical Reynolds number $Re_c$ as a function of
$\hat\mu$.  The dotted curve to the left of $\hat\mu=0.25$ is a purely
nonmagnetic instability.  The curves to the right of $\hat\mu=0.25$ are
the magneto-rotational instability.  Solid curves are $Pm=10^{-6}$,
dashed curves $Pm=10^{-5}$.  The numbers beside pairs of curves indicate
$\beta$.  Finally, the dot on the $\beta=4$ curves corresponds to the
solution shown in Fig.\ 2.}
\end{figure}
 
The explanation for this very different behavior for non-zero
$\beta$ lies in the coupling between the azimuthal field $b$ and the
meridional circulation $\psi$.  If $\beta=0$ these quantities are not
directly coupled at all, only indirectly through $v$ and $a$.  For
non-zero $\beta$ each acts directly on the other.  Of these two new
terms in the equations, the more important one turns out to be the
effect of $\psi$ on $b$.  Physically, this corresponds to the
meridional circulation $\psi$ advecting the original $B_\phi$ and
thereby generating a contribution to $b$.  It is this new way of
maintaining $b$ that allows this instability to proceed even in the
$Pm=0$ limit, where the classical $\beta=0$ MRI fails.
 
Figure 2 shows an example of these new solutions.  We note that
however dramatically $Re_c$ may have been reduced, the spatial
structure is much the same as for $\beta=0$, consisting of Taylor
vortices elongated slightly in the $z$-direction.  There is actually
one subtle difference, namely that the up/down symmetry in $z$ has
been broken.  This is due to the handedness of the imposed field,
which distinguishes between $\pm z$ in a way that a purely axial
field does not.  As a result of this symmetry-breaking, these new
modes are also no longer stationary, but instead have 
Im$(\gamma)\ne0$, corresponding to a drift in $z$, at the rate
Im$(\gamma)/k$.  This particular solution has Im$(\gamma)=0.153$ and
$k=2.33$, for a drift rate of $0.066\,(r_i\Omega_i)$.

\begin{figure}
\includegraphics{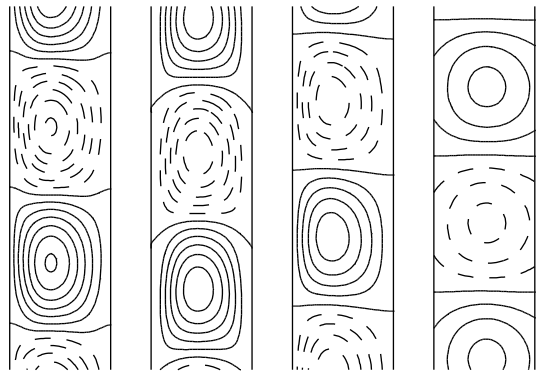}
\caption{The spatial structure of the eigenmode for $Pm=0$,
$\beta=4$ and $\hat\mu=0.27$, for which $Re_c=1521$ (see Fig.\ 1),
$H\!a=16.3$ and $k=2.33$.  From left to right, contours of $v$, $\psi$,
$b$ and $a$, with the contour interval for $b$ and $a$ 1/3 that of
$v$ and $\psi$.  Note the lack of any particular phase relationship
between the various quantities, due to the $\pm z$ symmetry-breaking.}
\end{figure}
 
Turning next to the Hartmann number $H\!a=16.3$, this translates into a
$\sim$12 G field, taking the fluid to be liquid sodium, and $r_i=10$ cm.
An axial field of that magnitude is certainly easily achievable in the
lab.  For the azimuthal field we then want 500 G cm/$r$, corresponding
to a current of 2500 A in this wire running down the central axis,
which is again achievable, if perhaps not quite so easily.
 
To summarize then, we have shown that the magneto-rotational
instability is radically altered if one imposes both axial and
azimuthal magnetic fields, becoming independent of the magnetic
Prandtl number in the $Pm\to0$ limit -- a result that is all the more
remarkable as neither purely axial nor purely azimuthal fields yield
anything like it.  With this new scaling, the critical Reynolds numbers
are reduced by several orders of magnitude, enough that these
instabilities could perhaps be attainable in the lab without being disrupted
by end-effects (although of course some end-effects will always be present,
particularly with this slow drift in $z$).  Further computational work
includes the nonlinear equilibration of these modes, as well as the
possibility of non-axisymmetric instabilities.  These and other issues
are currently being explored.
 
Finally, returning briefly to the original astrophysical motivation,
we note that virtually all astrophysical bodies do in fact have both
axial and azimuthal magnetic fields.  These magnetic fields are
typically not externally imposed though, but rather generated by electric
currents flowing within the system itself.  Self-consistently solving
for both the large-scale fields as well as the small-scale
instabilities is then far more complicated than our analysis here, but
the basic ingredients are certainly there for this new type of
magneto-rotational instability to play a role.

\begin{acknowledgments}
This work was developed during the `Magnetohydrodynamics of Stellar
Interiors' program at the Isaac Newton Institute for Mathematical Sciences.
We thank the Newton Institute and the program organizers for inviting us
both.
\end{acknowledgments}

\end{document}